**Detection of nontrivial topology driven by charge density wave in a semi-Dirac metal**

Rafiqul Alam[1#], Prasun Boyal[1#], Shubhankar Roy[2], Ratnadwip Singha[2], Buddhadeb Pal[1], Riju Pal[1], Prabhat Mandal[1,2], Priya Mahadevan[1*] and Atindra Nath Pal[1*]

[1]*Department of Condensed Matter and Materials Physics, S. N. Bose National Centre for Basic Sciences, Sector III, Block JD, Salt Lake, Kolkata 700106, India.*

[2]*Saha Institute of Nuclear Physics, Homi Bhabha National Institute, Kolkata 700 064, India*

# equal contribution

*priya@bose.res.in , atin@bose.res.in

**Abstract**

The presence of electron correlations in a system with topological order can lead to exotic ground states. Considering single crystals of $LaAgSb_2$ which has a square net crystal structure, one finds multiple charge density wave transitions (CDW) as the temperature is lowered. We find large planar Hall (PHE) signals in the CDW phase, which are still finite in the high temperature phase though they change sign. Optimising the structure within first-principles calculations, one finds an unusual chiral metallic phase. This is because as the temperature is lowered, the electrons on the Ag atoms get more localized, leading to stronger repulsions between electrons associated with atoms on different layers. This leads to successive layers sliding with respect to each other, thereby stabilising a chiral structure in which inversion symmetry is also broken. The large Berry curvature associated with the low temperature structure explains the low temperature PHE. At high temperature the PHE arises from the changes induced in the tilted Dirac cone in a magnetic field. Our work represents a route towards detecting and understanding the mechanism in a correlation driven topological transition through electron transport measurements, complemented by ab-initio electronic structure calculations.

**Introduction**

The past two decades have seen the experimental realization of a new class of materials, the topological materials[1–8]. The classification, entirely based on their electronic band structure and the underlying symmetry, has led to a set of universal physical properties independent of other details of the material[9],[10]. For instance, topological insulators have an energy gap in the bulk, while the surface states are protected by time-reversal symmetry[11]. This would in a two-dimensional example lead to the edge states being protected from back-scattering, and hence the possibility of dissipation less currents along the edge, while in three dimensions one would have topological magneto-electric effects[10],[12] with a quantized response. Another topological class - the Dirac or Weyl semimetals [13–19], involves materials with gapless bulk states which could host relativistic chiral fermions.

Various topological materials have been successfully identified in terms of their topological invariants using electronic structure calculations, which are effective single-particle approaches. However, more exotic effects are expected when electron-electron interactions need to be included[20–22]. There have been suggestions that attractive electron-electron interactions would drive a Weyl semimetal into a topological superconductor[23–26]. On the other hand, incommensurate charge density waves are predicted for those Weyl semi-metals in which one must include repulsive electron-electron interactions, with unusual axion electrodynamics being predicted in some of them[27–30], while several other emergent topological phases possessing novel quasiparticles such as quantum spin-hall insulator[31], fractional Chern insulator[32] or even the possibility of axial Higgs mode[33] have been predicted.

With an aim to identify candidate materials with these exotic ground states, in this work we examine the unusual electronic structure of $LaAgSb_2$. This is an example of a square net compound, a structure type in which several intermetallic compounds have been found to be topological semi-metals[34–36]. Apart from a tilted Dirac cone confirmed by ARPES measurements and first principle calculations, $LaAgSb_2$ is found to exhibit two CDW transitions, one at 211 K which is associated with an incommensurate wave over ~ 40 unit cells along the lattice vector **a** and another incommensurate CDW along the **c** direction taking place at a lower temperature of 186 K in which commensurate order involving 6 unit cells appears as the temperature is lowered[37]. However, the nature and origin of the CDW are still debated. While the Fermi surface measurements have revealed a nesting vector close to the calculated one, suggesting Fermi surface nesting as the origin of the CDW[38], there are a lot of experimental observations against a nesting driven CDW scenario. These observations are consistent with the growing understanding that while one may have regions of the Fermi surface parallel to each other, usually other effects take over and the direction of the CDW modulation has little or no connection with the nesting vector. NMR experiments reveal the emergence of *only* two distinct La species which seems unusual[39] despite the long wavelength associated with the CDW at 211 K. More importantly, the recent observation of coexisting

CDW and superconductivity below 300mK[40], further suggests the presence of a strong electron-phonon coupling as has been used to discuss the properties of the cuprates[41], NbSe$_2$ [42] and 2H-TaSe$_2$[43].

Electronically, another intriguing aspect is the observation of quasi-linear magnetoresistance (MR) with no saturation with distinct features above and below the CDW transitions[44]. The origin of unsaturated quasi-linear MR at low temperature is still elusive. For an isotropic Dirac cone, it has been simply attributed to the quantum limit of the possible Dirac fermions[45] at a relatively low field, while for systems with an anisotropic and tilted Dirac cone, it has been predicted to arise whenever there is nonvanishing Berry curvature, indicating the topological nature of the electronic bands[46–48]. Clearly, a better understanding of the coupling between the CDW phase and the topological properties is necessary.

Planar Hall effect (PHE), which is defined as the transverse voltage when the magnetic field and electric field are coplanar, has recently emerged as a transport-based probe to detect the non-trivial bands [49,50] in various topological semimetals or insulators such as ZrTe$_5$[51], Cd$_3$As$_2$[52] GdPtBi[53], VAl$_3$[54], Ta$_3$SiTe$_6$[55], Bi$_2$Se$_3$[56], and SmB$_6$[57]. In this letter, we report the detection of an unusual topological transition in single crystals of LaAgSb$_2$ mediated by CDW transitions through electronic transport measurements and first principle calculations. Experimentally we observe an anomaly in the zero-field resistivity and ordinary Hall coefficient at ~ 211 K, suggesting the reconstruction of the fermi surface mediated by the CDW transition. We also observe the quasi-linear magnetoresistance in the low temperature phase similar to previous reports[44]. However, the most intriguing result is the observation of PHE from low temperature to room temperature with a change of sign in the amplitude of the PHE above T$_{CDW1}$ (~ 211 K). Through our calculations, we observe a phonon mediated CDW phase in contrast to a Fermi-surface nesting driven mechanism. This leads to the unusual stabilization of a chiral metal phase at low temperatures, accompanied by a breaking of inversion symmetry which allows for a finite Berry curvature to be present, explaining the observed PHE. The origin of PHE in the high temperature phase is rather unusual and emerges due to the anisotropic nature of Dirac cone, which leads to an anisotropic response in a magnetic field, thereby explaining the observed PHE.

**Results:**

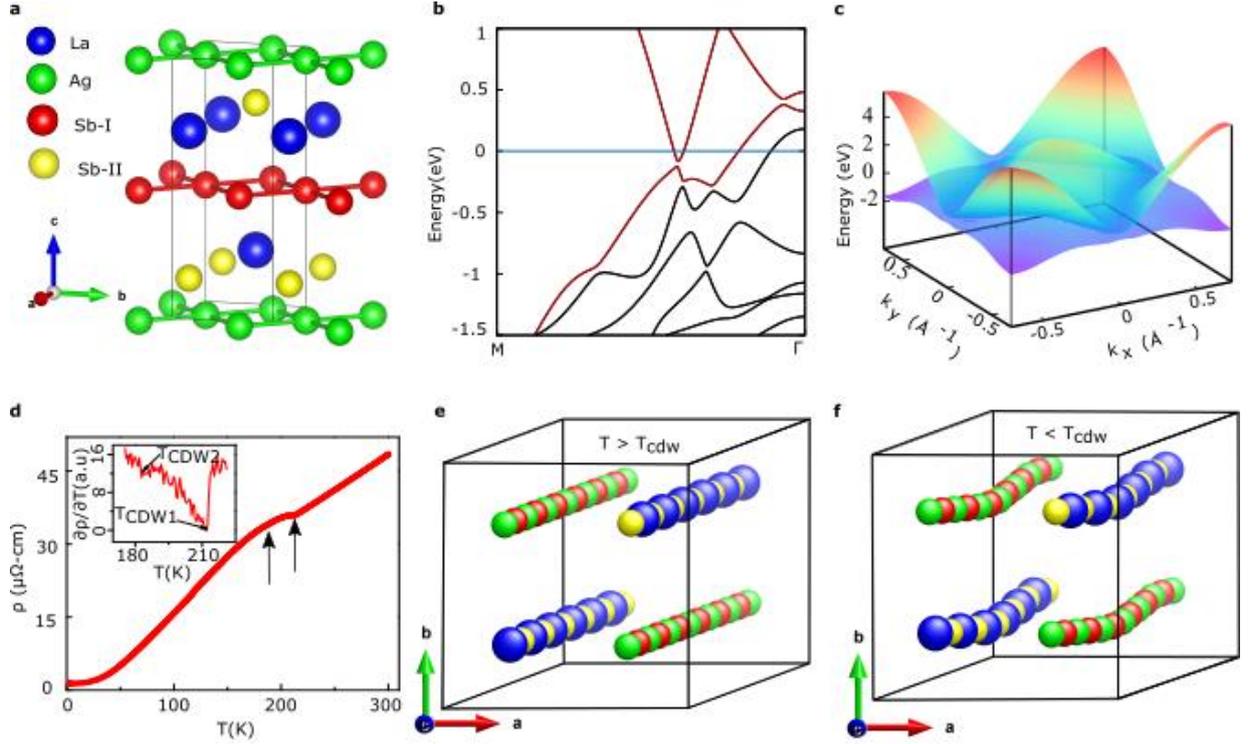

**Fig.1 a**. The room temperature tetragonal crystal structure of LaAgSb$_2$ (side view). **b.** The electronic band structure along the M-Γ direction with red bands showing the linearly dispersive bands (zero energy corresponds to the Fermi energy). **c.** Dispersion of the bands that crosses the Fermi energy across the $k_z = 0$ plane of the Brillouin zone. **d.** Resistivity (ρ) vs. temperature (T) plot exhibiting anomalies near the two CDW transitions indicated by the black arrows. Inset, showing the corresponding $\partial\rho/\partial T$ vs. $T$ curve to depict two CDW transitions ($T_{CDW1}$ and $T_{CDW2}$) more clearly. Top view of the crystal structure of LaAgSb$_2$ **(e)** without CDW modulation (T>$T_{CDW}$) and **(f)** with CDW modulation (T< $T_{CDW}$).

LaAgSb$_2$ crystallizes in the tetragonal *P4/nmm* space group. As shown in **Fig. 1a**, the material consists of two types of Sb atoms. One type of Sb atoms (red balls) as well as the Ag atoms (green balls) arrange themselves in a square net to form two-dimensional layers. These layers are stacked alternately along the **c** axis of the crystal. La ions (blue balls) and the other type of Sb (yellow balls) atoms are sandwiched within these layers. The corresponding Brillouin zone is shown in **Supplementary Fig. S1b** along with the high symmetry points. The density of states (DOS) at the Fermi level (See **Supplementary Fig. S1a**) is dominated by states from the two-dimensional Sb layers with small contributions from La d-states and the p-states of the other Sb atoms. The calculated electronic band structure reveals several linear bands crossing the Fermi energy along different high symmetry directions (**Supplementary Fig. S1c**). In **Fig. 1b** such crossing along the MΓ direction is shown. The Dirac cone has a gap of 40 meV, which agrees well with previous studies[38]. The anisotropic nature of the Dirac cone is evident in **Fig. 1c** where the dispersion of the bands, crossing Fermi energy, are shown in the 2D plane of the

Brillouin zone, defined by $k_z = 0$. Focussing next on the temperature dependent electronic resistivity, an anomaly is observed in the zero-field resistivity ($\rho$ vs. T plot in **Fig. 1d**) when the sample is cooled below a characteristic temperature ($T_{CDW1}$=211 K) similar to previous report[44]. There is another anomaly present in this system taking place at $T_{CDW2}$ which is clear from the $\frac{\partial \rho}{\partial T}$ vs. T plot (Inset of **Fig. 1d**). Both the transitions at $T_{CDW1}$=211 K and $T_{CDW2}$ =183 K are associated with the partial opening of a gap due to a CDW forming along the **a** and **c** axis, respectively[37]. The CDW transition taking place at 211 K has been associated with an incommensurate order appearing, involving around 40 unit cells along the lattice vector **a**. Consequently, one would expect the emergence of several inequivalent lattice sites. Surprisingly however, one finds just two distinct La sites emerging from NMR measurements[39]. Additionally, these measurements suggest the presence of an underlying periodic unit. The lower temperature CDW transition at 183 K which is also incommensurate develops into a commensurate transition at 164 K involving 6-unit cells in the **c** direction [37]. We focus on this transition in order to understand what is happening in the system.

As the temperature is lowered, the electrons on the Ag atoms get more localized. As the Sb atoms sit directly above the Ag atoms in the high temperature structure (**Fig. 1e**), the electrons on Sb experience a stronger Coulomb repulsion from the electrons on Ag. Consequently, on allowing for atomic displacements to minimize the total energy in a supercell involving 6 repeat units along the **c** lattice vector, one finds that atoms in neighbouring layers displace in the xy plane and form a modulated structure as shown in **Fig. 1f**. The optimised unit cell that we find is lower in energy by 11 meV per formula unit. While it is strange to find a chiral structure in a metallic system, LaAgSb$_2$ has very weak bonding in the **c** direction. This is supported by our calculations for the exfoliation energy which is found to be 115 meV/Å$^2$ (**Supplementary note 2**), placing it in the regime of potentially exfoliable materials[58]. Very low energy phonon modes have been found in our calculations (**Supplementary Figs. S2a, S2b**) as seen in other layered systems[59]. These modes involve shear modes allowing for the layer to slide as well as for layer compressions, which allow for the modulated structure to develop. This is consistent with experimental observations of low energy modes observed in reflectivity experiments[59] . The band dispersions, calculated for the modulated structure, have been unfolded onto the Brillouin zone of the high temperature structure. One finds a small gap opening in the RX direction (**Supplementary Fig. S4**), while the other directions remain gapless.

**Planar Hall effect and magnetoresistance measurement:**

We next focus on the effect of the CDW on the topological transport properties mediated by Dirac/Weyl Fermion by examining the planar Hall effect in these systems. In order to measure the planar Hall signal, we rotated the magnetic field in a way such that the current and magnetic field always remained in the same plane and measured the Hall resistivity and magnetoresistance (MR) at different angles between the current and magnetic field ($\theta$).

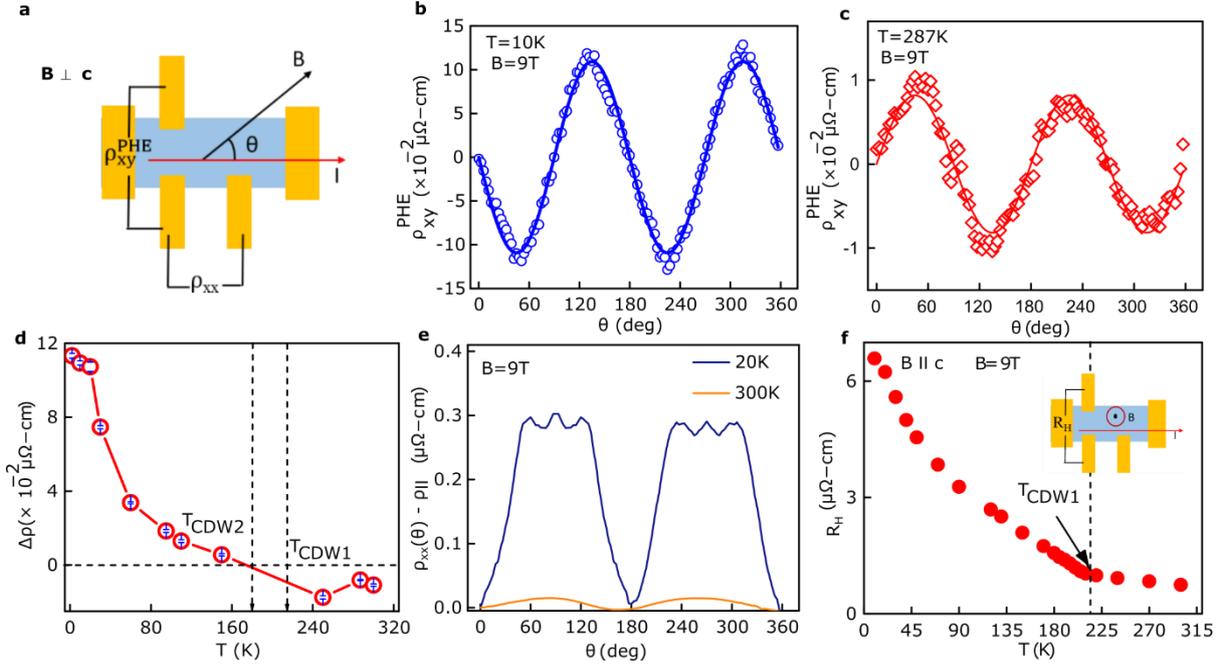

**Fig. 2| a.** Experimental configuration of PHE measurement set-up. Angular variation of PHE resistivity ($\rho_{xy}^{PHE}$) at **(b)** T=10K and **(c)** T=287K with the fitted curve (solid lines) by **Eq. 1**. **d.** Temperature dependence of $\Delta\rho = \rho_\perp - \rho_{||}$, extracted from the fitting of the PHE signal using **Eq. 1**. Both the $T_{CDWS}$ are indicated by vertical dotted lines. **e.** Angular variation of the in-plane resistivity difference ($\rho_{xx}(\theta) - \rho_{||}$) for B=9T at temperatures below and above $T_{CDWS}$ (T=20 K and 300 K). **f.** The temperature variation of the ordinary Hall coefficient ($R_H$), measured at B= 9T.

According to the general theory of anisotropic resistivity for isotropic topological metal[49], when the magnetic field is rotating in the sample plane, in-plane transverse (PHE) and longitudinal resistivity can be written respectively as,

$$\rho_{xy}^{PHE} = -\Delta\rho \sin\theta \cos\theta \quad (1)$$

$$(\rho_{xx} - \rho_{||}) = \Delta\rho \sin^2\theta \quad (2)$$

where, $\Delta\rho = \rho_\perp - \rho_{||}$ is the measure of resistivity anisotropy, $\rho_\perp$ and $\rho_{||}$ indicate the resistivity when the magnetic field is perpendicular and parallel to the electric current, respectively, and may have different forms depending on their origin. $\rho_{xy}^{PHE}$ and $\rho_{xx}$ both show a period of $180^0$. In specific, for $\rho_{xy}^{PHE}$, the valleys are at the angle $45^0$ and $225^0$ with peaks at $135^0$ and $315^0$, which is totally different from the angular dependence of ordinary Hall effect (OHE) as it holds a period of $360^0$. The device schematic is shown in **Fig. 2a**, where two longitudinal electrodes are used for in-plane longitudinal MR measurements and two transverse electrodes are used for obtaining the PHE signal. Possible errors arising from the misalignment have been considered (see **Supplementary note 4** for details).

After removing all the misalignment components, the intrinsic PHE signal for two temperatures, as representative, T=287 K and 10 K with B=9T are shown in **Figs. 2b** and **2c**. While the PHE signal shows the usual characteristic features with valleys at $45^0$ ($225^0$) and peaks at $135^0$ ($315^0$) in the CDW phase (at 10K), in the normal phase, it exhibits an $180^0$ phase-shift having peaks at $45^0$($225^0$) and valleys at $135^0$ ($315^0$). The PHE data is further fitted with **Eq. 1** and the extracted $\Delta\rho$ vs. T with a fixed magnetic field of 9T, is shown in **Fig. 2d**. It clearly shows that the PHE coefficient gradually decreases with an increase in temperature from 2 K to 110 K and becomes almost undetectable below ~150K. More interestingly, PHE coefficient starts to grow again in amplitude above $T_{CDW2}$ (211 K) and increases with increasing temperature, albeit, with a negative sign signifying $\rho_\perp < \rho_V$. Notably, a detectable PHE signals with $\Delta\rho = -ve$, is observed at room temperature of magnitude ~$1.08\times 10^{-2}$ μΩ cm which is one order less than that of T=2 K ($\Delta\rho = 1.13 \times 10^{-1}$μΩ-cm) value. The sign change of the PHE signal around the CDW transitions is reproducible and also verified in another sample (see **Supplementary Fig. S6**). Following this, we also investigated the impact of the CDW on the in-plane anisotropic magnetoresistance (AMR), which approximately follows **Eq. 2** at both below and above the CDW transitions (as shown in **Fig. 2e**). However, at certain angles, there are additional peaks similar to $KV_3Sb_5$[60] in the CDW phase which is invisible in the high temperature counterpart (above $T_{CDW}$=211 K). Furthermore, the temperature dependence of antisymmetric out-of-plane Hall coefficient (see inset, **Fig. 2f** for configuration of OHE measurement) shows an abrupt change near the CDW transitions (**Fig. 2f**), indicating a significant Fermi surface reconstruction due to the formation of CDWs. Magnetic field dependent Hall resistivity at several temperatures is demonstrated in **supplementary Fig. S7a**. Undoubtedly, both the temperature dependent planar Hall and planar resistivity signal together with the change in the ordinary Hall coefficient suggest a distinct change in the topological nature of the electronic bands mediated by the formation of charge density waves.

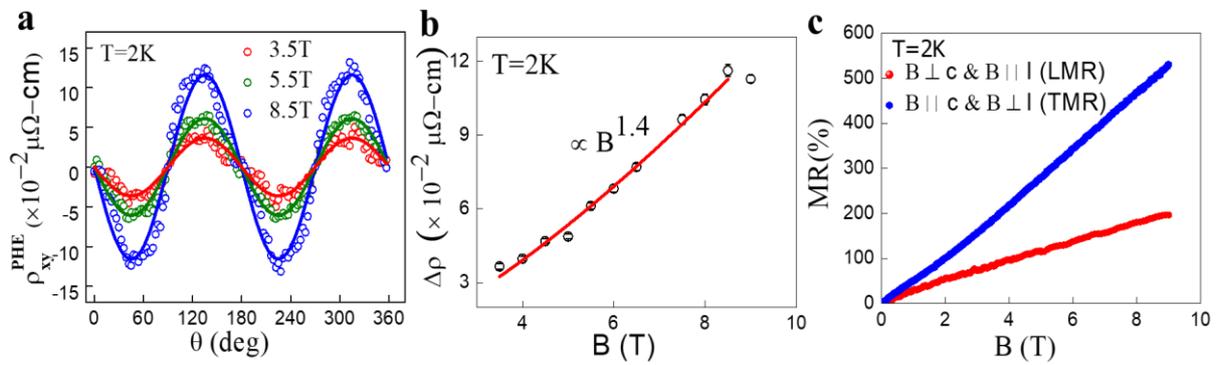

**Fig. 3| a.** Planar Hall resistivity ($\rho_{xy}^{PHE}$) of LaAgSb$_2$ as a function of angle between the current (I) and magnetic field (B) at different B (T=2 K). **b.** Magnetic field dependence of $\Delta\rho = \rho_\perp - \rho_\parallel$, extracted by fitting the PHE signals with **Eq. 1**. for different magnetic fields. The red solid line shows the fitted curve with a power law of $B^\alpha$, where $\alpha = 1.4$. **c.** Magnetic field evolution of longitudinal magnetoresistance (LMR) and transverse magnetoresistance (TMR) of the same sample with temperature fixed at 2 K.

To gain a further insight about the PHE, we measured the angular PHE signal in the CDW state (T=2 K) at different magnetic fields as shown in **Fig. 3a**. Notably, **Fig. 3b** depicts that the PHE amplitude, $\Delta\rho$ grows with magnetic field according to a power law $B^{1.4}$. We also measured the longitudinal MR (LMR) at T=2 K while I and B were kept parallel to each other, typically used to detect the chiral anomaly assisted negative MR[61,62]. However, we rather found a positive LMR which increases quasi-linearly with the applied magnetic field (see **Fig. 3c**), suggesting that additional mechanisms might be responsible for the observed magneto-transport behaviour. It is also noticed that the transverse MR (TMR), measured by applying a perpendicular magnetic field, increases quasi-linearly with B in the CDW phase (T=2 K), as shown in **Fig. 3c,** which is in line with the previous report[44]. It is also clear from the angle dependent TMR (see **Supplementary Fig. S7b**) that the Fermi surface is quasi-2D in nature.

**Theoretical understanding of PHE:**

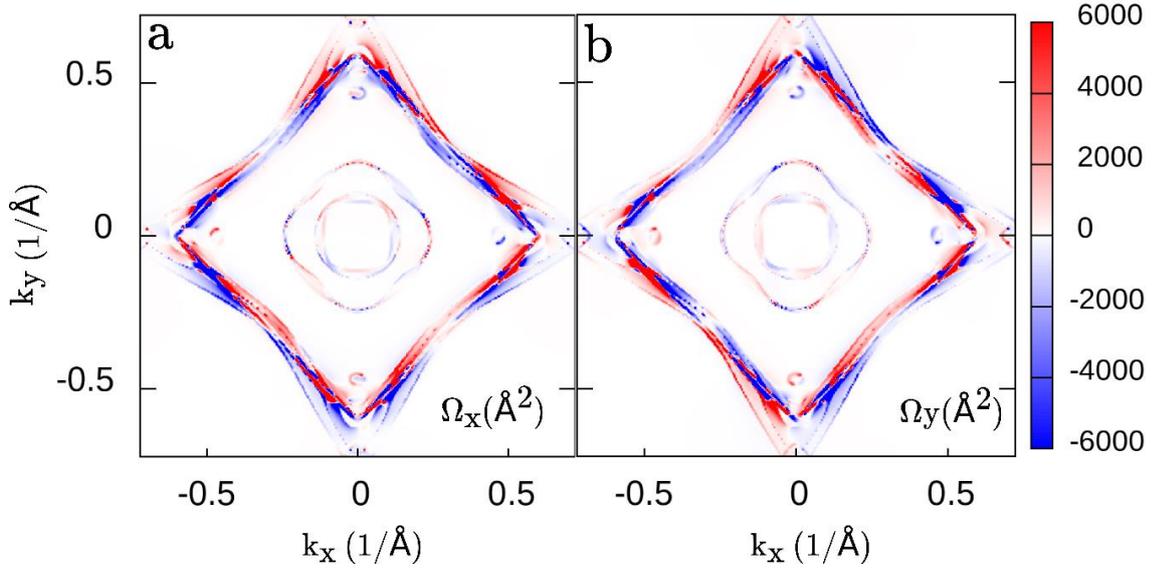

**Fig. 4|** 2D color plots of the **a.** x and **b.** y component of Berry curvature in $k_z=0$ plane in the CDW phase.

We will now attempt to understand the PHE signal and its unusual temperature dependent behaviour. The high temperature structure shown in **Fig. 1a** is centro-symmetric and the material is nonmagnetic. Hence, the system preserves both inversion and time reversal symmetry. Consequently, the intrinsic Berry curvature is zero indicating a non-topological origin in the PHE at high temperature. As the Dirac cones are anisotropic, this gives rise to an anisotropy in the in-plane resistivity in the presence of a magnetic field as discussed in the literature[63]. As back scattering between states at **k** and **-k** is allowed in the direction parallel to the in-plane magnetic field, while that is not the case in the direction perpendicular to the field, one finds that $\rho_\parallel$ becomes larger than $\rho_\perp$ which leads to a planar Hall signal with negative amplitude according to the **Eq. 1** [49].

The low temperature structure that we find (**Fig. 1f**) in the CDW phase breaks inversion symmetry. This leads to a large intrinsic Berry curvature whose x and y components are shown in **Figs. 4a** and **b**, respectively, consistent with the large amplitude of the PHE signal seen at low temperature. In such systems one also expects to observe negative LMR and a positive PHE amplitude ($\Delta\rho$) induced by chiral anomaly, both varying quadratically with B (in the low field limit). However, we observe a quasi-linear positive LMR along with a quasi-linear positive $\Delta\rho$. In addition to the nonvanishing Berry curvature, our theoretical calculations in the CDW phase clearly indicate that emergent Weyl cones are tilted. Hence, the observed LMR and PHE signals appear due to the combination of quadratic B term arising from the imbalance in the chiral chemical potential and a linear in B corrections to the resistivity induced by the tilted Weyl cones[46,47,64]. Furthermore, our findings of linear unsaturated positive TMR and the observed quasi-linear thermo-electric behaviour as reported in Ref.[44] can also be assigned to the same tilted Dirac cone driven mechanism[65].

In summary, we have considered $LaAgSb_2$, a material in which square net structures are present, and examined the evolution of its electronic structure with temperature and the consequent topological properties. While tilted Dirac cones have been associated with its high temperature structure, the presence of inversion symmetry suggests that it would not have topological properties. This makes the observation of planar Hall effect both at high temperature as well as at low temperature puzzling, coupled with the fact that there is a sign change and enhancement in its value below the CDW transitions. As the structure of the system is not known in the CDW phases, we used geometry optimisations within first-principles electronic structure calculations to determine the low temperature structure. In spite of the three-dimensional network of the compound, we find a small exfoliation energy placing it in the regime of an exfoliable two-dimensional materials, with very low energy shear phonon modes. This is consistent with the primary distortion found in the optimised crystal structure in which neighbouring layers are displaced with respect to each other, leading to a chiral structure which is surprisingly metallic too. The increased localization of the electrons on the Ag atoms as the temperature is lowered, increases the Coulomb repulsions between electrons on atoms belonging to atoms in different layers, driving this shear instability. This at low temperature breaks the inversion symmetry and consequently has a large Berry curvature associated with it. The planar Hall effect in the high temperature phase can be traced to the tilted Dirac cone in a magnetic field.

**Experimental details:**

Single crystals of $LaAgSb_2$ were made by the Sb-flux technique [65,66]. Elemental La (Alfa Aesar 99.9%), Ag (Alfa Aesar 99.99%), and Sb (Alfa Aesar 99.999 9%) in molar ratio 1: 2:20 were kept in an alumina crucible. The crucible was then placed in a quartz tube and sealed under dynamic vacuum. The quartz tube was kept at temperature 1050 °C for 12 h, and slowly cooled (2 °C/h) to 670 °C . At this temperature, the crystals were isolated from the flux of Sb by centrifugation. The obtained crystal is rectangular with crystallographic *c* axis along the perpendicular of the plane. The Energy Dispersive X-Ray Spectroscopy (EDX) was done for elemental analysis on different regions of the single crystal. EDX color mapping along with elemental ratio is provided in the **Supplementary Figure S8.** The transport measurements were performed in a physical property measurement system (Quantum Design, USA) using the ac-transport option. Conducting silver paste and gold wires were attached to make the electrical contacts in a four-probe configuration.

**Theoretical calculation details**:

The *ab-initio* electronic structure is calculated using projector augmented-wave method[67,68] implementation of density functional theory (DFT) within the Vienna Ab-initio Simulation Package (VASP)[69–71]. The atom positions are optimized to reduce the forces acting on the atom to lie below the tolerance value of 10 meV/Å. Perdew-Burke-Ernzerhof potentials[72,73] was used for the exchange correlation functional. For the self-consistent calculation, a k-space grid of 8x8x4 was used for the unit cell. The cut-off energy for the plane wave basis was taken to be 600 eV. The effective tight binding model is obtained by constructing maximally localized Wannier functions (MLWF) implemented in WANNIER90[71]. The basis consists of *d* orbitals of La and Ag and *p* orbitals of Sb atoms. Berry Curvature is calculated using the WANNIER TOOLS[74] package.

**Acknowledgements:**

RA acknowledges DST-Inspire (award number IF170926) and PB acknowledges CSIR for fellowship. P.M. acknowledges SERB for funding through SPF/2021/000066 and DST through DST-INT/SWD/VR/P-08/2019. The support and the resources provided by PARAM Shivay Facility under the National Supercomputing Mission, Government of India at the Indian Institute of Technology, Varanasi are gratefully acknowledged. ANP appreciates support from DST-Nano Mission (grant no. DST/NM/TUE/QM-10/2019). This research has made use of the characterization facilities under the Technical Cell and Technical Research Centre (TRC) of S. N. Bose National Centre for Basic Sciences.

**References:**

[1]   F. D. M. Haldane, *Physical Review Letters* **1988**, *61*, 2015.

[2]   C. L. Kane, E. J. Mele, *Physical Review Letters* **2005**, *95*, 1.

[3]   L. Fu, C. L. Kane, *Physical Review B - Condensed Matter and Materials Physics* **2007**, *76*, 1.

[4]   L. Fu, C. L. Kane, E. J. Mele, *Physical Review Letters* **2007**, *98*, 1.

[5]   D. Hsieh, D. Qian, L. Wray, Y. Xia, Y. S. Hor, R. J. Cava, M. Z. Hasan, *Nature* **2008**, *452*, 970.

[6]   T. Topological, B. Te, Y. L. Chen, J. G. Analytis, J. Chu, Z. K. Liu, S. Mo, X. L. Qi, H. J. Zhang, D. H. Lu, X. Dai, Z. Fang, S. C. Zhang, I. R. Fisher, Z. Hussain, Z. Shen, **2009**, *178*, 1.

[7]   Y. Zhang, K. He, C. Z. Chang, C. L. Song, L. L. Wang, X. Chen, J. F. Jia, Z. Fang, X. Dai, W. Y. Shan, S. Q. Shen, Q. Niu, X. L. Qi, S. C. Zhang, X. C. Ma, Q. K. Xue, *Nature Physics* **2010**, *6*, 584.

[8]   J. E. Moore, L. Balents, *Physical Review B - Condensed Matter and Materials Physics* **2007**, *75*, 3.

[9]   M. Z. Hasan, C. L. Kane, *Reviews of Modern Physics* **2010**, *82*, 3045.

[10]  X. L. Qi, S. C. Zhang, *Reviews of Modern Physics* **2011**, *83*.

[11]  J. E. Moore, *Nature* **2010**, *464*, 194.

[12]  X. L. Qi, T. L. Hughes, S. C. Zhang, *Physical Review B - Condensed Matter and Materials Physics* **2008**, *78*, 1.

[13]  S. Murakami, *New Journal of Physics* **2007**, *9*.

[14]  G. B. Halász, L. Balents, *Physical Review B - Condensed Matter and Materials Physics* **2012**, *85*, 1.

[15]  S. M. Huang, S. Y. Xu, I. Belopolski, C. C. Lee, G. Chang, B. Wang, N. Alidoust, G. Bian, M. Neupane, C. Zhang, S. Jia, A. Bansil, H. Lin, M. Z. Hasan, *Nature Communications* **2015**, *6*.

[16]  B. Q. Lv, H. M. Weng, B. B. Fu, X. P. Wang, H. Miao, J. Ma, P. Richard, X. C. Huang, L. X. Zhao, G. F. Chen, Z. Fang, X. Dai, T. Qian, H. Ding, *Physical Review X* **2015**, *5*, 1.

[17]  B. Q. Lv, N. Xu, H. M. Weng, J. Z. Ma, P. Richard, X. C. Huang, L. X. Zhao, G. F. Chen, C. E. Matt, F. Bisti, V. N. Strocov, J. Mesot, Z. Fang, X. Dai, T. Qian, M. Shi, H. Ding, *Nature Physics* **2015**, *11*, 724.

[18]     H. Weng, C. Fang, Z. Fang, B. Andrei Bernevig, X. Dai, *Physical Review X* **2015**, *5*, 1.

[19]     S. Y. Xu, I. Belopolski, D. S. Sanchez, C. Zhang, G. Chang, C. Guo, G. Bian, Z. Yuan, H. Lu, T. R. Chang, P. P. Shibayev, M. L. Prokopovych, N. Alidoust, H. Zheng, C. C. Lee, S. M. Huang, R. Sankar, F. Chou, C. H. Hsu, H. T. Jeng, A. Bansil, T. Neupert, V. N. Strocov, H. Lin, S. Jia, M. Zahid Hasan, *Science Advances* **2015**, *1*.

[20]     A. Vishwanath, T. Senthil, *Phys Rev X* **2013**, *3*, 1.

[21]     L. Fidkowski, X. Chen, A. Vishwanath, *Phys Rev X* **2014**, *3*, 1.

[22]     C. Wang, A. C. Potter, T. Senthil, *Phys Rev B Condens Matter Mater Phys* **2013**, *88*, 1.

[23]     Y. Li, F. D. M. Haldane, *Phys Rev Lett* **2018**, *120*, 1.

[24]     X. L. Qi, T. L. Hughes, S. C. Zhang, *Phys Rev B Condens Matter Mater Phys* **2010**, *81*, 1.

[25]     Y. Li, C. Wu, *Sci Rep* **2012**, *2*, 1.

[26]     S. Han, C. S. Tang, L. Li, Y. Liu, H. Liu, J. Gou, J. Wu, D. Zhou, P. Yang, C. Diao, J. Ji, J. Bao, L. Zhang, M. Zhao, M. V. Milošević, Y. Guo, L. Tian, M. B. H. Breese, G. Cao, C. Cai, A. T. S. Wee, X. Yin, *Advanced Materials* **2023**, *35*.

[27]     Z. Wang, S. C. Zhang, *Phys Rev B Condens Matter Mater Phys* **2013**, *87*, 1.

[28]     Y. You, G. Y. Cho, T. L. Hughes, *Phys Rev B* **2016**, *94*, 1.

[29]     J. Gooth, B. Bradlyn, S. Honnali, C. Schindler, N. Kumar, J. Noky, Y. Qi, C. Shekhar, Y. Sun, Z. Wang, B. A. Bernevig, C. Felser, *Nature* **2019**, *575*, 315.

[30]     K. Zhang, N. Zou, Y. Ren, J. Wu, C. Si, W. Duan, *Adv Funct Mater* **2022**, *32*.

[31]     J. L. Xiaofeng Qian, Junwei Liu, Liang Fu, **2014**, *346*, 1344.

[32]     H. Polshyn, Y. Zhang, M. A. Kumar, T. Soejima, P. Ledwith, K. Watanabe, T. Taniguchi, A. Vishwanath, M. P. Zaletel, A. F. Young, *Nature Physics* **2022**, *18*, 42.

[33]     Y. Wang, I. Petrides, G. McNamara, M. M. Hosen, S. Lei, Y. C. Wu, J. L. Hart, H. Lv, J. Yan, D. Xiao, J. J. Cha, P. Narang, L. M. Schoop, K. S. Burch, *Nature* **2022**, *606*, 896.

[34]     S. Klemenz, S. Lei, L. M. Schoop, *Annu Rev Mater Res* **2019**, *49*, 185.

[35]     R. Singha, A. K. Pariari, B. Satpati, P. Mandal, *Proc Natl Acad Sci U S A* **2017**, *114*, 2468.

[36]     S. Lei, S. M. L. Teicher, A. Topp, K. Cai, J. Lin, G. Cheng, T. H. Salters, F. Rodolakis, J. L. McChesney, S. Lapidus, N. Yao, M. Krivenkov, D. Marchenko, A. Varykhalov, C. R. Ast, R. Car, J. Cano, M. G. Vergniory, N. P. Ong, L. M. Schoop, *Advanced Materials* **2021**, *33*.

[37] C. Song, J. Park, J. Koo, K. B. Lee, Y. Rhee, L. Bud'ko, C. Canfield, N. Harmon, I. Goldman, *Physical Review B - Condensed Matter and Materials Physics* **2003**, *68*, 1.

[38] X. Shi, P. Richard, K. Wang, M. Liu, C. E. Matt, N. Xu, R. S. Dhaka, Z. Ristic, T. Qian, Y. F. Yang, C. Petrovic, M. Shi, H. Ding, *Physical Review B* **2016**, *93*, 1.

[39] S. H. Baek, S. L. Bud'ko, P. C. Canfield, F. Borsa, B. J. Suh, *Physical Review B* **2022**, *106*, 1.

[40] K. Akiba, N. Umeshita, T. C. Kobayashi, *Physical Review B* **2022**, *106*, 1.

[41] J. M. Tranquada, *Neutron News* **1996**, *7*, 17.

[42] K. Cho, M. Kończykowski, S. Teknowijoyo, M. A. Tanatar, J. Guss, P. B. Gartin, J. M. Wilde, A. Kreyssig, R. J. McQueeney, A. I. Goldman, V. Mishra, P. J. Hirschfeld, R. Prozorov, *Nature Communications* **2018**, *9*, 1.

[43] S. H. Baek, Y. Sur, K. H. Kim, M. Vojta, B. Büchner, *New Journal of Physics* **2022**, *24*, 2.

[44] K. Wang, C. Petrovic, *Physical Review B - Condensed Matter and Materials Physics* **2012**, *86*, 1.

[45] A. Abrikosov, *Physical Review B - Condensed Matter and Materials Physics* **1998**, *58*, 2788.

[46] K. Das, A. Agarwal, *Physical Review B* **2019**, *99*, 1.

[47] A. Kundu, Z. Bin Siu, H. Yang, M. B. A. Jalil, *New Journal of Physics* **2020**, *22*, 2.

[48] D. Xiao, J. Shi, Q. Niu, *Physical Review Letters* **2005**, *95*, 1.

[49] A. A. Burkov, *Physical Review B* **2017**, *96*, 1.

[50] S. Nandy, G. Sharma, A. Taraphder, S. Tewari, *Physical Review Letters* **2017**, *119*, 1.

[51] P. Li, C. H. Zhang, J. W. Zhang, Y. Wen, X. X. Zhang, *Giant planar Hall effect in the Dirac semimetal $ZrTe_{5-\delta}$*, Vol. 98, **2018**, p. 121108(R).

[52] H. Li, H. W. Wang, H. He, J. Wang, S. Q. Shen, *Physical Review B* **2018**, *97*, 1.

[53] N. Kumar, S. N. Guin, C. Felser, C. Shekhar, *Physical Review B* **2018**, *98*, 1.

[54] R. Singha, S. Roy, A. Pariari, B. Satpati, P. Mandal, *Physical Review B* **2018**, *98*, 1.

[55] S. Roy, R. Singha, A. Ghosh, P. Mandal, *Physical Review Materials* **2021**, *5*, 3.

[56] Y. Wang, S. V. Mambakkam, Y. X. Huang, Y. Wang, Y. Ji, C. Xiao, S. A. Yang, S. A. Law, J. Q. Xiao, *Physical Review B* **2022**, *106*, 1.

[57] L. Zhou, B. C. Ye, H. B. Gan, J. Y. Tang, P. B. Chen, Z. Z. Du, Y. Tian, S. Z. Deng, G. P. Guo, H. Z. Lu, F. Liu, H. T. He, *Physical Review B* **2019**, *99*, 1.

[58] N. Mounet, M. Gibertini, P. Schwaller, D. Campi, A. Merkys, A. Marrazzo, T. Sohier, I. E. Castelli, A. Cepellotti, G. Pizzi, N. Marzari, *Nature Nanotechnology* **2018**, *13*, 246.

[59]    K. H. Michel, B. Verberck, *Physica Status Solidi (B) Basic Research* **2012**, *249*, 2604.

[60]    L. Li, E. Yi, B. Wang, G. Yu, B. Shen, Z. Yan, M. Wang, *npj Quantum Materials* **2023**, *8*, 1.

[61]    X. Huang, L. Zhao, Y. Long, P. Wang, D. Chen, Z. Yang, H. Liang, M. Xue, H. Weng, Z. Fang, X. Dai, G. Chen, *Physical Review X* **2015**, *5*, 1.

[62]    H. Li, H. He, H. Z. Lu, H. Zhang, H. Liu, R. Ma, Z. Fan, S. Q. Shen, J. Wang, *Nature Communications* **2016**, *7*, 1.

[63]    S. H. Zheng, H. J. Duan, J. K. Wang, J. Y. Li, M. X. Deng, R. Q. Wang, *Physical Review B* **2020**, *101*, 1.

[64]    D. Ma, H. Jiang, H. Liu, X. C. Xie, *Physical Review B* **2019**, *99*, 1.

[65]    B. Jiang, J. Zhao, J. Qian, S. Zhang, X. Qiang, L. Wang, R. Bi, J. Fan, H. Z. Lu, E. Liu, X. Wu, *Physical Review Letters* **2022**, *129*, 56601.

[66]    L. Gondek, B. Penc, A. Szytula, N. Stusser, *Journal of Alloys and Compounds* **2002**, *346*, 80.

[67]    P. E. Blöchl, *Physical Review B* **1994**, *50*, 17953.

[68]    D. Joubert, *Physical Review B - Condensed Matter and Materials Physics* **1999**, *59*, 1758.

[69]    G. Kresse, J. Hafner, *Physical Review B* **1993**, *47*, 558.

[70]    G. Kresse, J. Furthmüller, *Computational Materials Science* **1996**, *6*, 15.

[71]    A. A. Mostofi, J. R. Yates, G. Pizzi, Y. S. Lee, I. Souza, D. Vanderbilt, N. Marzari, *Computer Physics Communications* **2014**, *185*, 2309.

[72]    J. P. Perdew, K. Burke, M. Ernzerhof, *Physical Review Letters* **1996**, *77*, 3865.

[73]    J. P. Perdew, K. Burke, M. Ernzerhof, *Physical Review Letters* **1997**, *78*, 1396.

[74]    Q. S. Wu, S. N. Zhang, H. F. Song, M. Troyer, A. A. Soluyanov, *Computer Physics Communications* **2018**, *224*, 405.

Supplementary Information

**Detection of nontrivial topology driven by charge density wave in a semi-Dirac metal**

Rafiqul Alam[1#], Prasun Boyal[1#], Shubhankar Roy[2], Ratnadwip Singha[2], Buddhadeb Pal[1], Riju Pal[1], Prabhat Mandal[1,2], Priya Mahadevan[1*] and Atindra Nath Pal[1*]

[1]Department of Condensed Matter and Materials Physics, S. N. Bose National Centre for Basic Sciences, Sector III, Block JD, Salt Lake, Kolkata 700106, India.

[2]Saha Institute of Nuclear Physics, Homi Bhabha National Institute, Kolkata 700 064, India

# equal contribution

*priya@bose.res.in , atin@bose.res.in

**Contents:**

**Supplementary note 1:** Electronic band structure, density of states of $LaAgSb_2$ in the Normal phase (room temperature).

**Supplementary note 2:** Additional information on layered structure of $LaAgSb_2$.

**Supplementary note 3:** Electronic band structure of $LaAgSb_2$ in the CDW phase (low temperature)

**Supplementary note 4:** Extraction of pure planar Hall signal from raw data.

**Supplementary note 5:** Planar Hall data from a second sample.

**Supplementary note 6:** Magnetic field dependence of ordinary Hall effect and angular magnetoresistance (AMR).

**Supplementary note 7:** Characterisation of single crystal using EDX and color mapping.

# 1. Electronic band structure, density of states of LaAgSb$_2$ in the Normal phase (room temperature).

The Brillouin zone of the high temperature structure as well as the atom projected partial density of states are shown below. Here, Sb-I corresponds to the atoms that belong to the square net formed, while and Sb-II represents the other Sb atoms.

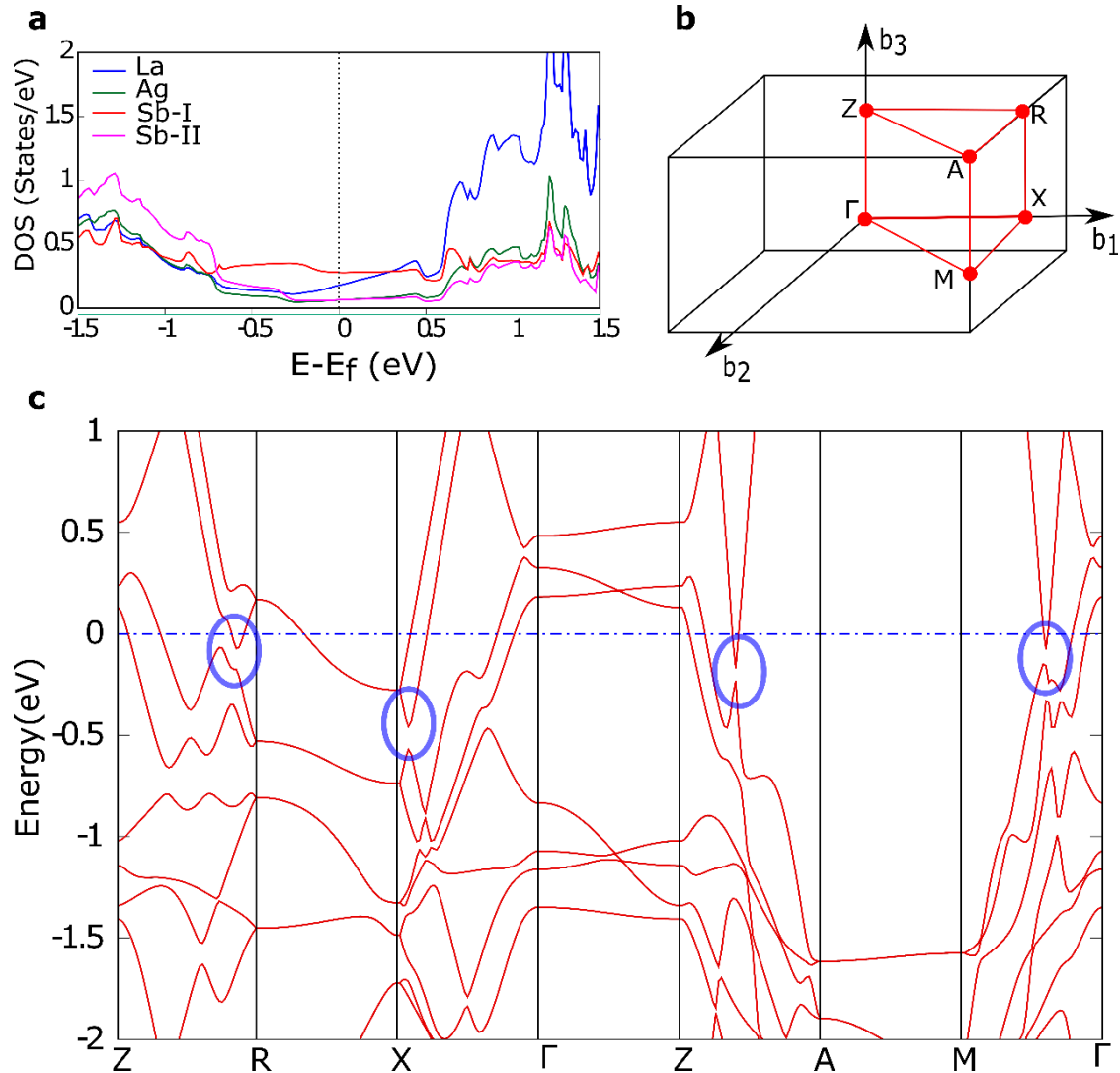

**Fig. S1: a.** Atom projected partial density of states. **b**. Brillouin zone of LaAgSb$_2$ **c.** Electronic band structure of LaAgSb$_2$ including spin orbit interactions. The zero of the energy is Fermi energy (indicated by the blue dashed line). The linear band dispersions in the vicinity of the fermi level have been indicated by the blue circled regions.

## 2. Additional information on layered structure of LaAgSb$_2$:

**Calculation of exfoliation energy:**

The exfoliation energy is a measure of how easily layers can be exfoliated from bulk structure. For LaAgSb$_2$ we have theoretically calculated the exfoliation energy using the following method. First two structures, infinitely periodic in a and b direction and truncated in c direction, are constructed. Truncation in the c direction is made such that one has only one layer and the other has two layers. The cleave plane is considered to be in between La atoms and the Sb square network. 20 Å vacuum is added in the c direction to ensure that the interactions between images of periodic supercells imposed by the method we use are negligible. Calculating the energy of the two structures, the exfoliation energy per unit area is,

$$E_{exfoliation} = \frac{2*E_{monolayer} - E_{bilayer}}{Area} \quad (S1)$$

**Low frequency phonon modes:**
The phonon spectra of the constructed layered structure are calculated using the force constants with finite displacement method as implemented in the PHONOPY code[1,2]. Supercells of 3 × 3 × 1-unit cells of the planar structure with Γ- centred mesh of 8 × 8 × 1 k points have been used to calculate force constants accurately. The low frequency sheer mode and compression mode for bilayer is shown in Fig. S2.

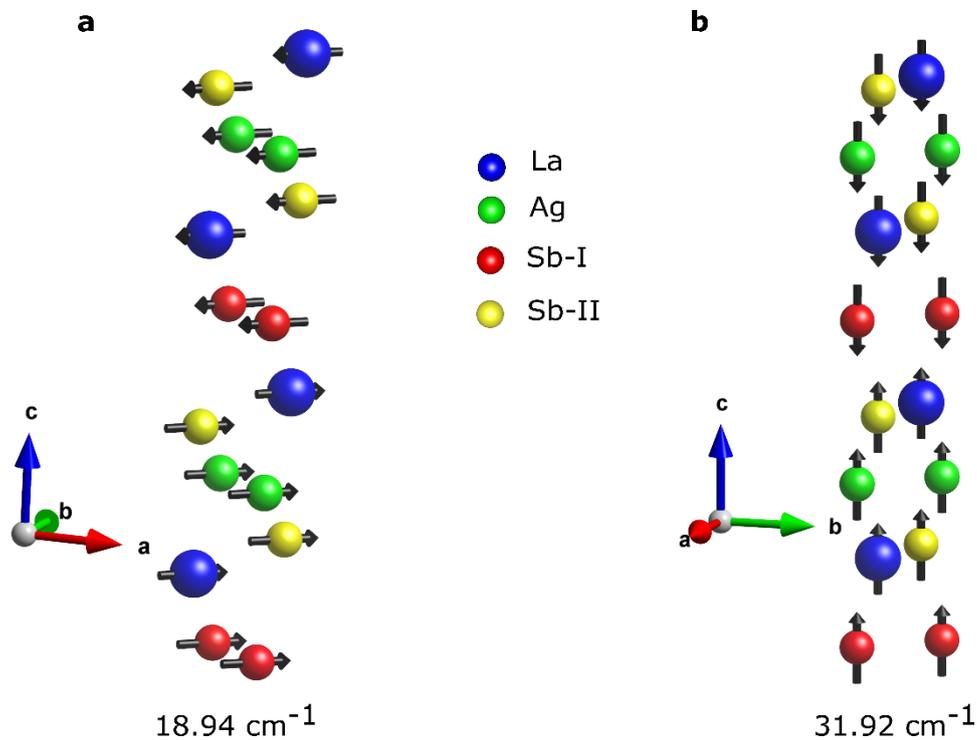

**Fig. S2|** The low frequency phonon modes for a bilayer with their frequencies indicated have been given. Panel **a.** shows a shear mode, while panel **b.** represents a compression mode.

**Two Types of La atom in CDW phase:**
Analysing the electrostatic potential of the La atoms in the CDW structure reveals two distinct groups of La atoms in the unit cell. This agrees with NMR results of LaAgSb$_2$, done in previous study[3]. In the following figure these two groups of La atoms are shown.

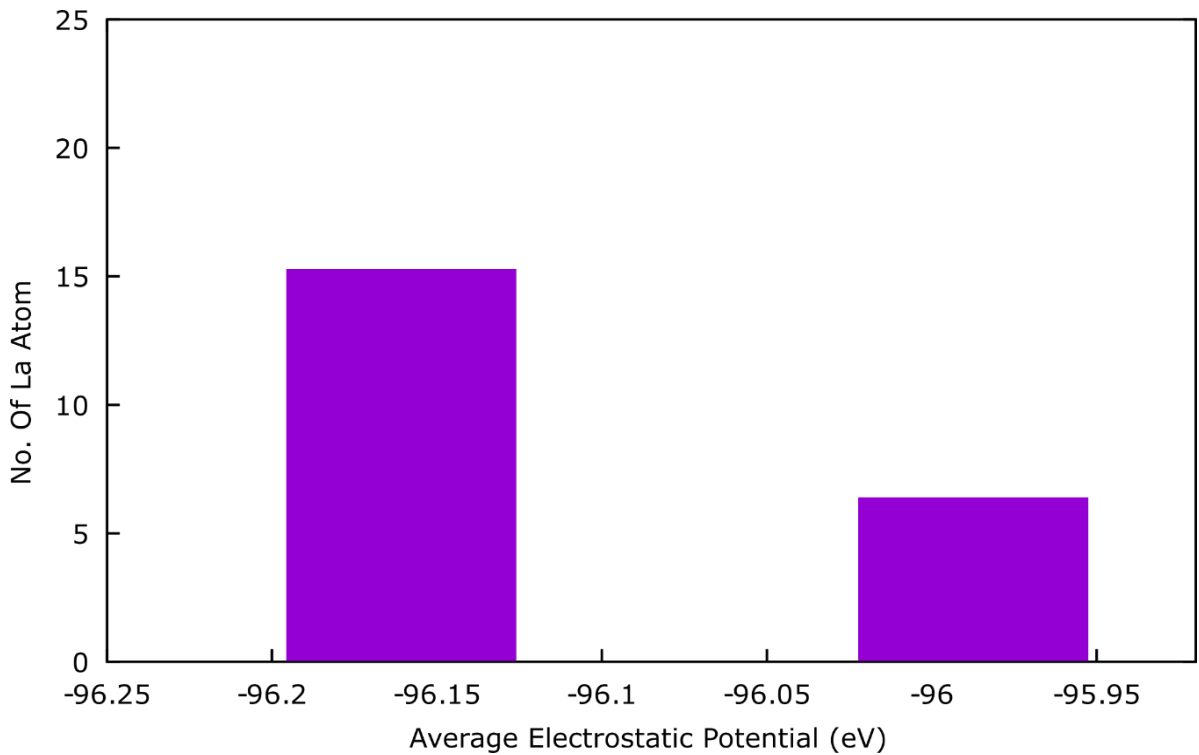

**Fig. S3|** Histogram plot of no. Of La atoms in the CDW structure as a function of their average electrostatic potential. Two distinct type of La atoms in the modulated structure is observed.

## 3. Electronic band structure of LaAgSb$_2$ in the CDW phase (low temperature)

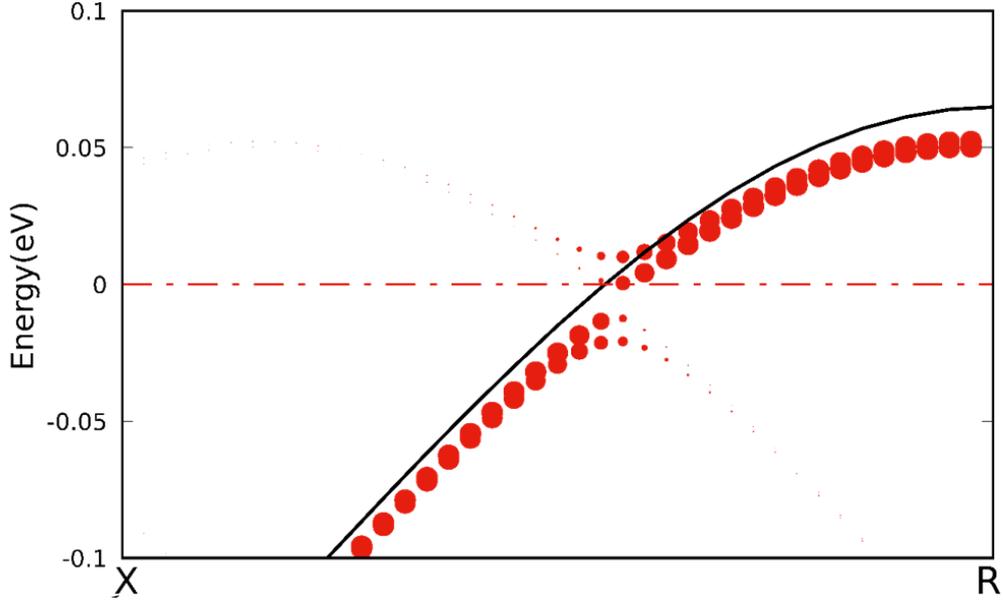

**Fig. S4|** Band structure of the modulated supercell unfolded onto the high temperature structure X to R direction. A small band gap opens because of the modulation in the c direction. Size of the red dots are proportional to the contribution of the supercell states at those k points. The black line corresponds to the band dispersion for the high temperature structure. The red dot-dashed line represents the Fermi energy in the calculation.

## 4. Extraction of pure planar Hall signal from raw data:

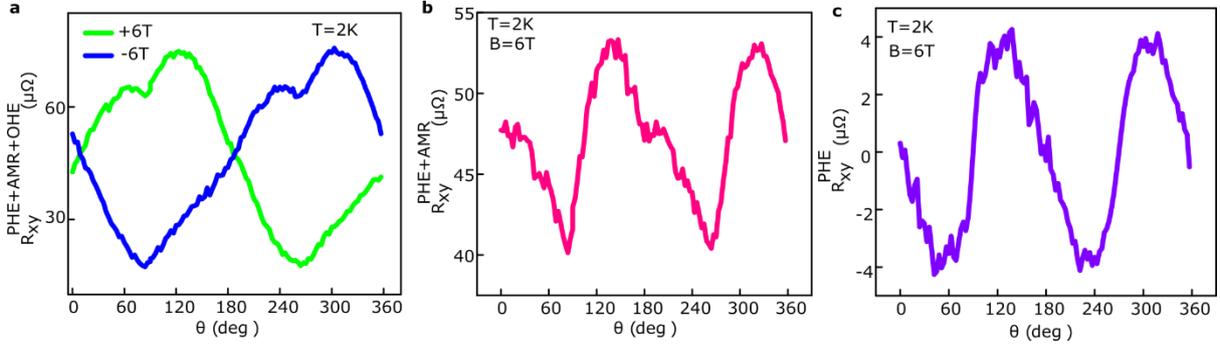

**Fig. S5|** **a.** $R_{xy}^{PHE+AMR+OHE}(+6T)$ and $R_{xy}^{PHE+AMR+OHE}(-6T)$ vs. θ at T=2K. **b.** $R_{xy}^{PHE+AMR}(+6T)$ vs. θ at T=2K after removing the out of plane Hall components by following the **Eq. S2**. **c.** $R_{xy}^{PHE}(+6T)$ vs. θ at T=2K after removing the angular MR part by following the **Eq. S3**.

Firstly, the rotating plane of the magnetic field is not always the same as the sample plane and a residual out of plane Hall voltage may be present during the planar measurement. The antisymmetric Hall effect and symmetric planar Hall effect behave differently under reversal of magnetic field direction with respect to the sample plane. The planar hall resistivity is the average of the resistivity with positive and negative fields. The formula for calculating the PHE signal is shown in the following equation[4]:

$$R_{xy}^{PHE+AMR} = \frac{[R_{xy}(B,\theta)+R_{xy}(-B,\theta)]}{2}  \quad (S2)$$

Secondly, the two electrodes for Hall measurement are not always perfectly aligned longitudinally, which further corrupts the planar Hall signal with an admixture of angular magnetoresistance. The actual PHE data can be recovered by processing the data using the formula.

$$R_{xy}^{PHE} = \frac{(R_{xy}(\theta)-R_{xy}(\pi-\theta))}{2} \quad (S3)$$

## 5. Planar Hall signal's data from a second sample

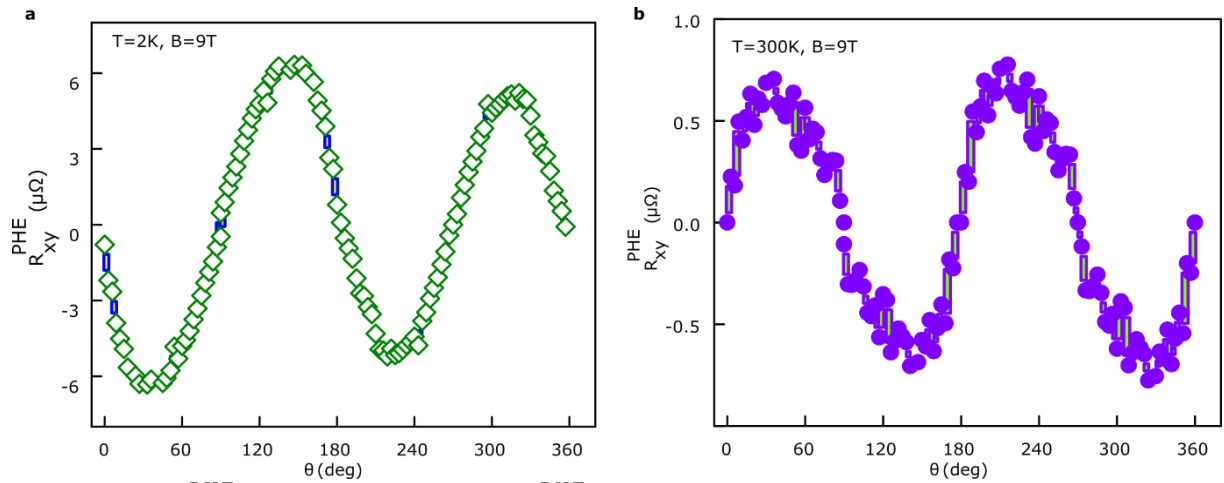

**Fig. S6| a.** $R_{xy}^{PHE}$(9T) vs. θ at T=2 K **b.** $R_{xy}^{PHE}$(9T) vs. θ at T=300 K.

The PHE signal displays its typical characteristics, including valleys at $45^0$ ($225^0$) and peaks at in the CDW phase (at 10 K), but in the normal phase (T=300 K), it displays a $180^0$ phase-shift with peaks at $45^0$($225^0$) and valleys at $135^0$ ($315^0$) in the sample 2.

## 6. Magnetic field dependence of ordinary Hall effect and angular magnetoresistance (AMR)

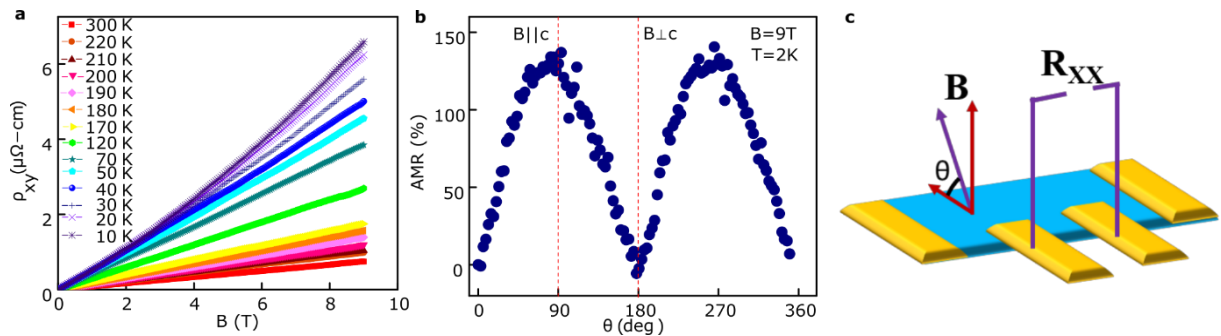

**Fig. S7| a.** The magnetic field dependence of ordinary Hall resistivity at several temperatures. **b**. Angular Magnetoresistance (AMR) vs. θ at T=2 K and B=9T. **c.** The schematic of the Angle dependent MR configuration.

As displayed in **Fig. S7a**, the Hall resistivity increases non linearly with increase of magnetic field(B) below T=50 K while it is linear in B above this temperature. This nonlinear Hall resistivity may be attributed to the multiband effect[5]. To provide evidence for a quasi-2d Fermi surface, we present the angle-dependent TMR in **Fig. S7b. When B is parallel to the c axis, the MR is large, while it is low when B lies in the ab plane (B ⊥ I) (the schematic of the measurement configuration is displayed in Fig. S7c). This result indicates the quasi-2d nature of the Fermi surface**[6].

**7. Characterisation of single crystal using EDX and color mapping:**

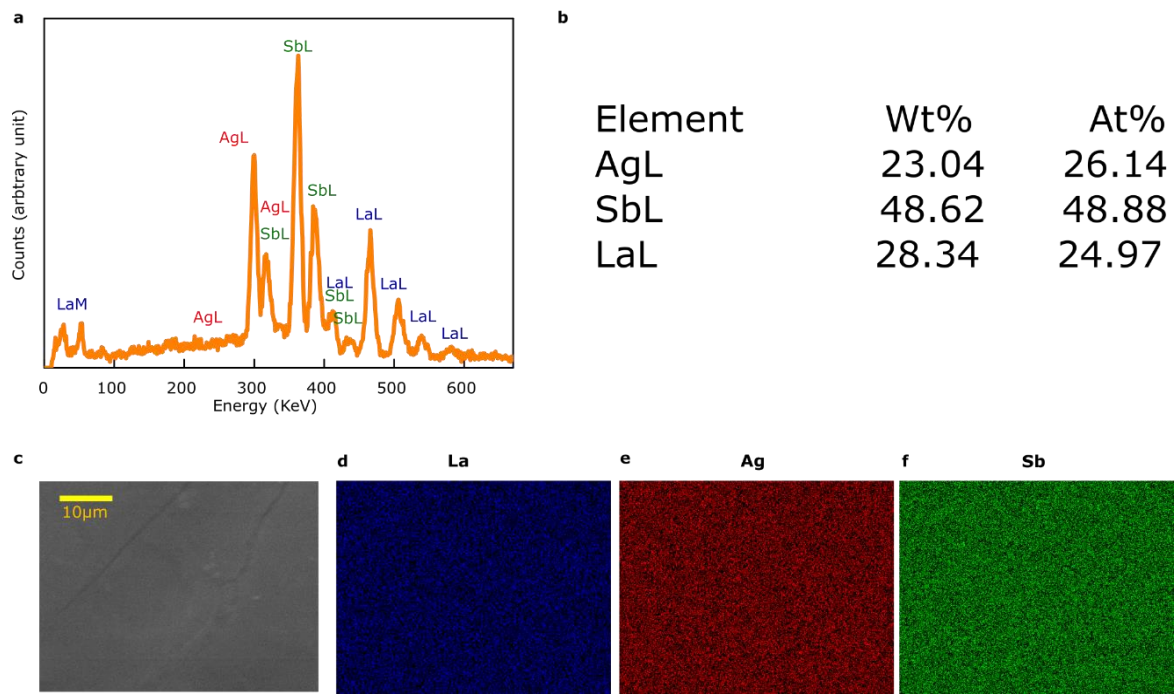

**Fig. S8| a.** The EDAX/ EDX spectroscopy data on a randomly selected area of a grown single crystal of LaAgSb$_2$. **b.** The elemental analysis of the EDAX data (**Fig. S8| a**.) for the elements present in the grown single crystal (where, Wt% and At% imply molar weight ratio and atomic weight ratio, respectively). The almost perfect At% with La: Ag: Sb≈1:1:2 suggests the good quality of the single crystal of LaAgSb$_2$. **c.** FESEM image of a grown single crystal of LaAgSb$_2$. **(d, e, f).** Overlay of elemental mappings of La, Ag and Sb.

**References:**

[1]    A. Togo, I. Tanaka, *Scripta Materialia* **2015**, *108*, 1.

[2]    A. Togo, *Journal of the Physical Society of Japan* **2023**, *92*.

[3]    S. H. Baek, S. L. Bud'ko, P. C. Canfield, F. Borsa, B. J. Suh, *Phys Rev B* **2022**, *106*, 1.

[4]    H. Li, H. W. Wang, H. He, J. Wang, S. Q. Shen, *Phys Rev B* **2018**, *97*, 1.

[5]    K. Wang, C. Petrovic, *Phys Rev B Condens Matter Mater Phys* **2012**, *86*, 1.

[6]    K. Wang, D. Graf, C. Petrovic, *Physical Review B - Condensed Matter and Materials Physics* **2013**, *87*, 1.